\documentclass[12pt]{iopart}

\usepackage{graphicx}
\usepackage{amsfonts}
\usepackage{epsfig}
\usepackage{dcolumn}
\usepackage{bm}
\usepackage{times}
\usepackage{xcolor}

\begin{document}

\title[]{Dynamics of social contagions with local trend imitation}

\author{Xuzhen Zhu$^{1}$, Wei Wang$^{2,3*}$, Shimin Cai$^{3,4,5}$,
  H. Eugene Stanley$^{5}$}

\address{$^{1}$ State Key Laboratory of Networking and Switching Technology,
Beijing University of Posts and Telecommunications, Beijing 100876, China}

\address{$^{2}$ College of Computer Science and Technology,
Chongqing University of Posts and
Telecommunications, Chongqing 400065, China}

\address{$^{3}$ Big Data Research Center, University of Electronic
  Science and Technology of China,
Chengdu 611731, China}

\address{$^{4}$ Web Sciences Center, School of Computer Science and
  Engineering, University of Electronic Science and Technology of China,
  Chengdu 611731, China}

\address{$^{5}$ Center for Polymer Studies and Department of Physics,
  Boston University, Boston, Massachusetts 02215, USA}

\ead{wwzqbx@hotmail.com}

\begin{abstract}
Research on social contagion dynamics has not yet including a
theoretical analysis of the ubiquitous local trend imitation (LTI)
characteristic. We propose a social contagion model with a tent-like
adoption probability distribution to investigate the effect of this LTI
characteristic on behavior spreading. We also propose a generalized
edge-based compartmental theory to describe the proposed model. Through
extensive numerical simulations and theoretical analyses, we find a
crossover in the phase transition: when the LTI capacity is strong, the
growth of the final behavior adoption size exhibits a second-order phase
transition. When the LTI capacity is weak, we see a first-order phase
transition. For a given behavioral information transmission probability,
there is an optimal LTI capacity that maximizes the final behavior
adoption size. Finally we find that the above phenomena are not
qualitatively affected by the heterogeneous degree distribution.  Our
suggested theory agrees with the simulation results.
\end{abstract}

\maketitle

\section{Introduction}
The study of social contagion has attracted wide attention among
researchers in the field of network science
\cite{Watts2007,Castellano2009}. Studies of social contagion have
focused on such subjects as behavior spreading \cite{Centola2011},
information spreading \cite{Gao2016Effective}, and the contagion of
sentiment \cite{Christakis2007}, and they have been both theoretical and
experimental in their exploration of the essential nature of social
contagion \cite{Christakis2007,Barrat2008}. Unlike biological contagions
(e.g., epidemic spreading)
\cite{Pastor-Satorras2001,Wang2014Asymmetrically,Shu2016Recovery},
social contagions have a reinforcement effect \cite{Porter2014}. A
useful approach to studying social contagions that includes the
reinforcement effect is a threshold model
\cite{Watts2002,Gleeson2007,Dodds2009,Gleeson2008} that assumes a
susceptible individual accepts a new behavior when a fraction
\cite{Watts2002} or number \cite{Granovetter1973} of its neighbors
greater than an adoption threshold already exhibit the behavior. This
threshold model is a trivial Markovian process.  Numerical simulations
and theoretical analyses verify that the social reinforcement effect can
alter the phase transitions of social contagions \cite{Watts2002}. In
particular, the final adoption size first grows continually and then
decreases discontinually versus the average degree.  Many non-Markovian
social contagion models have also been developed to depict the social
reinforcement effect
\cite{Dodds2004,Zheng2013,Wang2015,Wang2015Dynamics,
  Wang2016Dynamics,Liu2017Social}. Recent research has found that social
reinforcement originates in the memory of non-redundant information
transmission \cite{Wang2015,Wang2015Dynamics,Wang2016Dynamics}, that the
growth of the final behavior adoption size is dependent on the
behavioral information transmission probability, and it changes from
continuous to discontinuous when the dynamical or structural parameters
are altered.

In real-world cases, the probability that an individual will adopt a new
behavior may be either positively or negatively correlated with the
number of neighbors who have already adopted the behavior. For example,
some style-conscious people who imitate the behavior of celebrities and
adopt the latest fashions may also strive to avoid anything that has
become overly-popular and ubiquitous (Leibenstein calls this the ``snob
effect'' \cite{Leibenstein1976Beyond}). Another example is when an
individual habitually patronizes a restaurant with good food and a
convivial atmosphere, but then avoids it when it becomes overly-popular
and crowded. Both of these examples exhibit the local trend imitation
(LTI) phenomenon \cite{Simmel1986Threshold,
  Granovetter1986Threshold,Dodds2013Limited}, i.e., the adoption
probability first increases with an increase in the number or fraction
of adopted neighbors and then decreases. Dodds et al.
 found the LTI effect induces the emergence
of chaos in Markovian social contagions \cite{Dodds2013Limited}.

Because the LTI effect in non-Markovian social contagions has not been
systematically analyzed, we here propose a social contagion model that
uses the LTI characteristic effect to describe the dynamics of behavior
spreading. The LTI characteristic effect is described using a tent-like adoption
probability distribution.  We develop a generalized edge-based
compartmental theory for quantitative validation. Both the numerical
simulation and theoretical results show that the LTI characteristic
strongly affects the final adoption size. In particular, when the LTI is
strong the system undergoes a discontinuous first-order phase
transition. When it is weak the system undergoes a continuous
second-order phase transition. For each spreading probability there is
an optimal LTI capacity that maximizes the final adoption size. We also
find that the heterogeneity level of the degree distribution does not
qualitatively affect the outcome.

\begin{figure}
\begin{center}
\includegraphics[height=10cm,width=12cm]{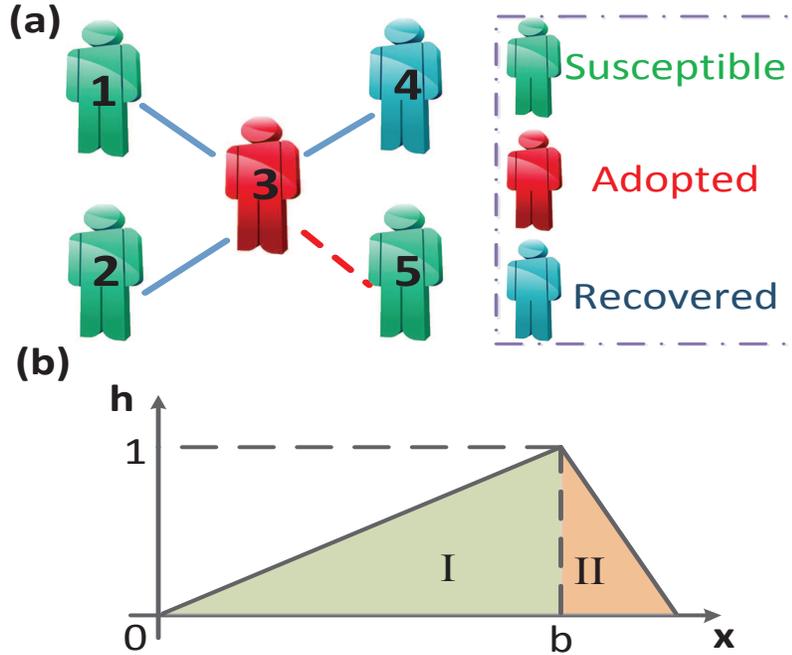}
\caption{(a) Illustration of social contagions on complex networks.  (b)
  Tent-like behavior adoption probability. Notations $b$ and $x$ indicate local trend imitation capacity and the ratio of adopted informants, respectively. In region I, the adoption probability increases with $x$. In region II, the adoption probability decreases with $x$.}
\label{illu}
\end{center}
\end{figure}

\section{Model Description}\label{Model}
We here use a generalized susceptible-adopted-recovered (SAR) model
\cite{Wang2015,Wang2015Dynamics,Wang2016Dynamics} to describe behavior
spreading in complex networks with $N$ nodes and a degree distribution
$P(k)$.  Figure~1(a) shows that at any given time each individual is in
either a susceptible (S), adopted (A) or recovered (R) state. An
individual in the susceptible state has not adopted the behavior. An
individual in the adopted state adopts the behavior and exhibits or
transmits it to susceptible neighbors. An individual in the recovered
state abandons the behavior and no longer exhibits or transmits it.

To include the LTI effect in social contagions, we use a tent-like
function $h(x,b)$ as the behavior adoption probability, defined as
\begin{equation} \label{h_x_b}
h(x,b) = \left\{ {\begin{array}{*{20}{c}}
{\frac{x}{b}, }&{~0 < x \le b },\\
{\frac{{1 - x}}{{1 - b}}, }&{~b < x < 1,}
\end{array}} \right.
\end{equation}
where $x$ is the ratio between an individual's received information and
its degree. The parameter $b$ is the LTI capacity of an
individual. When $0<x\leq b$, i.e., region I in Fig.~\ref{illu}(b), the
adoption probability increases with $x$. Thus region I is the
\emph{promotion region}. When $b<x<1$, i.e., region II in
Fig.~\ref{illu}(b), the adoption probability decreases with $x$.  Region
II is the \emph{depression region}. Small $b$ increases the LTI
capacity, and when the value of $b$ is large, the LTI capacity decreases.

We initially randomly select a seed to be an adopter and allow the rest
to remain susceptible. At each time step, every adopted individual $v$
transmits behavioral information to every susceptible neighbor with a
probability $\lambda$.  If a susceptible neighbor $u$ of $v$ receives
the information, the cumulative pieces of information $m$ collected by
$u$ increases by one, i.e., $m=m+1$. We disallow multi-transmission of information
between individuals $u$ and $v$, i.e., only non-redundant
information transmission is allowed. Individual $u$ becomes adopted with
a probability $h(m/k,b)$, in which $k$ is the degree of individual
$u$. Thus the system is non-Markovian.  Every adopted individual 
abandons the behavior with a probability $\gamma$, and moves to the
recovered state. The spreading dynamics terminate when all adopted
individuals have moved to the recovered state.

\section{Theoretical analysis}\label{theory}
To describe our proposed model, we use
Refs.~\cite{Wang2015,Miller2011,Miller2013} and develop a generalized
edge-based compartmental theory.  We define mathematical symbols $S(t)$,
$A(t)$, and $R(t)$ as the fraction of individuals in the susceptible,
adopted, and recovered states at time step $t$, respectively.

We assume that individual $u$ in the cavity state \cite{Karra2010}
receives behavioral information from adopted neighbors but does not
transmit it further.  We define $\theta(t)$ to be the probability that
individual $v$ by time $t$ has not transmitted the behavioral
information to individual $u$ along a randomly selected edge.  By time
$t$, an individual $u$ with degree $k$ has received $m$ pieces of
behavioral information from different neighbors at probability
\begin{equation} \label{active_a_k}
\phi_m(k,t)= {k \choose m}[\theta(t)]^{k-m}[1-\theta(t)]^m.
\end{equation}

Individual $u$, with degree $k$ and $m$ units of received information,
remains susceptible with a probability $\prod\limits_{j = 0}^m {[1 -
    h(\frac{j}{k},b)]}$.  We determine the probability that individual
$u$ with degree $k$ has received $m$ units of information and by time
$t$ is still in susceptible state with
\begin{eqnarray} \label{S_k_t}
\fl S(k,t) &= \sum\limits_{m = 0}^k {{\phi _m}(k,t)\prod\limits_{j =
    0}^m {[1 - h(\frac{j}{k},b)]} } \nonumber\\
\fl  &= \sum\limits_{m = 0}^{\left\lfloor {bk} \right\rfloor } {{\phi
    _m}} (k,t)\prod\limits_{j = 0}^m {[1 - \frac{j}{{bk}}]} {\rm{ }} +
\sum\limits_{m = \left\lceil {bk} \right\rceil }^k {{\phi _m}}
(k,t)\prod\limits_{j = 0}^{\left\lceil {bk} \right\rceil } {[1 -
    \frac{j}{{bk}}]} \prod\limits_{j = \left\lceil {bk} \right\rceil }^m
           {[1 - \frac{{1 - \frac{j}{k}}}{{1 - b}}]}.
\end{eqnarray}
Considering all possible degrees $k$, we calculate the total ratio of
susceptible individuals to be
\begin{equation} \label{S_t}
S(t) = \sum\limits_k {P(k)S(k,t)}.
\end{equation}
Neighbor $v$ of individual $u$ is either susceptible, adopted, or
recovered, thus $\theta(t)$ can be divided, i.e.,
\begin{equation} \label{theta_t}
\theta (t) = {\xi_S}(t) + {\xi_A}(t) + {\xi_R}(t),
\end{equation}
where ${\xi_S}(t)$ [${\xi_A}(t)$, ${\xi_R}(t)$] is the probability that
neighbor $v$ of individual $u$ is in the susceptible (adopted,
recovered) state and has not transmitted the behavioral information to
$u$ by time $t$.

When individual $v$ with degree $k'$ is initially susceptible, they
cannot transmit behavioral information to $u$, but can receive
information from all $k'-1$ neighbors except susceptible $u$.  Thus we
determine the probability that individual $v$ by time $t$ has received
$m$ units of information to be
\begin{equation} \label{tau_m}
\phi_m(k'-1,t)={k^\prime-1 \choose m}
[\theta(t)]^{k^\prime-m-1}[1-\theta(t)]^m.
\end{equation}
Taking into consideratin all possible values of $m$, we determine
the probability that individual $v$ with degree $k^\prime$ remains
susceptible to be
\begin{footnotesize}
\begin{eqnarray} \label{big_theta_k_t}
\fl \Theta (k',t) &= \sum\limits_{m = 0}^{k' - 1} {{\phi _m}}
(k'-1,t)\prod\limits_{j = 0}^m {[1 - h(\frac{j}{{k'}},b)]} \nonumber\\
\fl  &= \sum\limits_{m = 0}^{\left\lfloor {bk'} \right\rfloor } {{\phi
    _m}} (k'-1,t)\prod\limits_{j = 0}^m {[1 - \frac{j}{{bk'}}]} {\rm{ }}
+
 \sum\limits_{m = \left\lceil {bk'} \right\rceil }^{k' - 1} {{\phi _m}}
 (k'-1,t)\prod\limits_{j = 0}^{\left\lceil {bk'} \right\rceil } {[1 -
     \frac{j}{{bk'}}]} \prod\limits_{j = \left\lceil {bk'} \right\rceil
 }^m {[1 - \frac{{1 - \frac{j}{{k'}}}}{{1 - b}}]}.
\end{eqnarray}
\end{footnotesize}
In an uncorrelated network, an edge connects an individual of degree
$k^\prime$ with probability $k^\prime P(k^\prime) /\langle k\rangle$,
where $\langle k\rangle$ is the average degree. We obtain
\begin{equation} \label{xi_S_t}
{\xi _S}(t) = \sum\limits_{k'} {\frac{{k'P(k')}}{{\langle k\rangle }}}
\Theta (k',t).
\end{equation}
If an adopted individual transmits behavioral information through an edge
with probability $\lambda$, $\theta(t)$ does not fulfill the definition,
and the decrease of the fraction of $\theta(t)$ equals
$\lambda{\xi_A}(t)$, which is
\begin{equation} \label{d_theta_t}
\frac{{d\theta (t)}}{{dt}} =  - \lambda {\xi _A}(t).
\end{equation}
If an adopted individual does not transmit the behavioral information
through any edge with probability $1-\lambda$ but moves into the
recovered state with probability $\gamma$, ${\xi _R}(t)$ will
consequently increase. We thus obtain
\begin{equation} \label{d_xi_R_t}
\frac{{d{\xi _R}(t)}}{{dt}} = \gamma (1 - \lambda ){\xi _A}(t).
\end{equation}
Using Eqs.~(\ref{d_theta_t}) and~(\ref{d_xi_R_t}), and the initial
conditions of $\theta(0)=1$ and ${\xi_R}(0)=0$, we get
\begin{equation} \label{xi_R_t}
{\xi _R}(t) = \frac{{\gamma [1 - \theta (t)](1 - \lambda )}}{\lambda }.
\end{equation}
Substituting ${\xi _S}(t)$, ${\xi _A}(t)$ and ${\xi _R}(t)$ of
Eq.~(\ref{theta_t}) into Eqs.~(\ref{xi_S_t}), (\ref{d_theta_t}), and
(\ref{xi_R_t}), respectively, we find the time evolution of $\theta(t)$
to be
\begin{equation} \label{d_theta_t_end}
\frac{{d\theta (t)}}{{dt}} =  - \lambda [\theta (t) - \sum\limits_{k'}
  {\frac{{k'P(k')}}{{\langle k\rangle }}\Theta (k',t)} ] + \gamma [1 -
  \theta (t)](1 - \lambda ).
\end{equation}

At each time step $t$, some susceptible individuals adopt the behavior
and some adopted individuals move into the recovered state. Note that the
growth of $A(t)$ is equivalent to the decrease of $S(t)$ minus the
fraction of adopted individuals that with probability $\gamma$ enter the
recovered state. Thus the time evolution of $A(t)$ is
\begin{eqnarray}\label{d_A_t}
\frac{{dA(t)}}{{dt}} &=  - \frac{{dS(t)}}{{dt}} - \gamma A(t) \nonumber\\
&= - \sum\limits_k {P(k)\frac{{dS(k,t)}}{{dt}}}  - \gamma A(t),
\end{eqnarray}
where
\begin{equation}\label{d_S_k t_2}
\fl \frac{{dS(k,t)}}{{dt}} = \sum\limits_{m = 0}^{\left\lfloor {bk}
  \right\rfloor } {\Psi (t)} \prod\limits_{j = 0}^m {[1 -
    \frac{j}{{bk}}]} {\rm{ }} + \sum\limits_{m = \left\lceil {bk}
  \right\rceil }^k {\Psi (t)} \prod\limits_{j = 0}^{\left\lceil {bk}
  \right\rceil } {[1 - \frac{j}{{bk}}]} \prod\limits_{j = \left\lceil
  {bk} \right\rceil }^m {[1 - \frac{{1 - \frac{j}{k}}}{{1 - b}}]},
\end{equation}
and
\begin{eqnarray}\label{d_phi_m_dt}
\fl \Psi (t) &= \frac{{d{\phi _m}(k,t)}}{{dt}} \nonumber\\
\fl &= \left( {\begin{array}{*{20}{c}}
k\\
m
\end{array}} \right)\left\{ {(k' - m - 1){{[\theta (t)]}^{k' - m -
      2}}{{[1 - \theta (t)]}^m} - m{{[\theta (t)]}^{k' - 1 - m}}{{[1 -
        \theta (t)]}^m}} \right\}.
\end{eqnarray}
The time evolution of $R(t)$ is
\begin{equation} \label{d_R_t}
\frac{{dR(t)}}{{dt}} = \gamma A(t).
\end{equation}

Equations~(\ref{active_a_k})--(\ref{S_t}) and
(\ref{d_theta_t_end})--(\ref{d_A_t}) describe social contagion in terms
of LTI, and they can be used to compute the fraction of each state at
any arbitrary time step.  When $t \to \infty$, we find the final
adoption size $R(\infty)$.

In the final state, we find that
\begin{equation} \label{theta_infty}
\theta (\infty ) = \sum\limits_{k^\prime}{\frac{{k^\prime P(k^\prime)}}
{{\langle k\rangle }}} \Theta (k^\prime,\infty ) + \frac{{\gamma [1 -
\theta (\infty )](1 - \lambda )}}{\lambda }.
\end{equation}
Note that $\theta(t)$ decreases with $t$ when adopted individuals
continually transmit the behavioral information to neighbors.  Thus when
there is more than one stable fixed point in Eq.~(\ref{theta_infty})
only the maximum stable fixed point is physically meaningful. Inserting
this value into Eqs.~(\ref{active_a_k})--(\ref{S_t}) gives us the steady
value of the susceptible density $S(\infty)$ and the final behavior
adoption size $R(\infty)$.

\begin{figure}[t]
\begin{center}
\includegraphics[height=8cm,width=16cm]{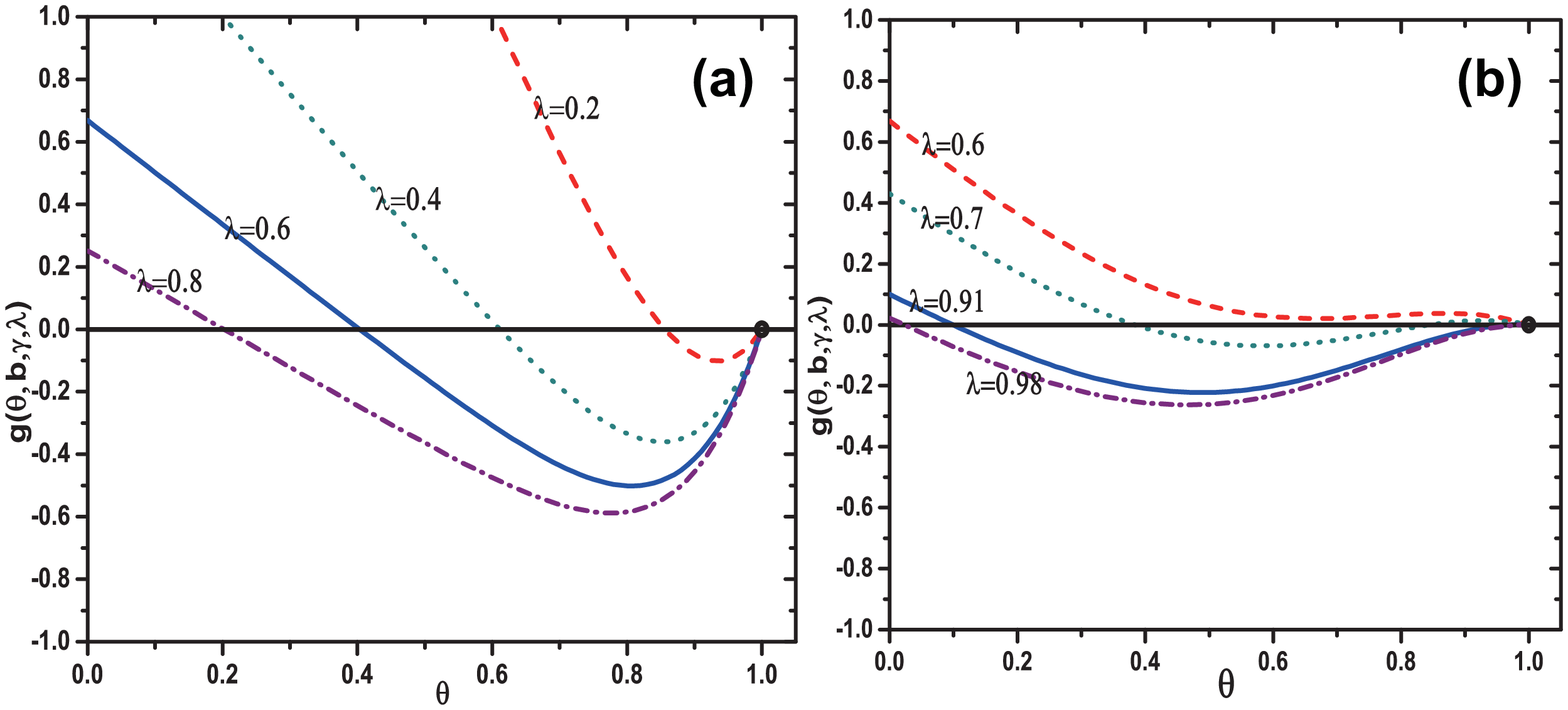}
\caption{(Color online) Demonstration of graphical solutions of
  Eq.~(\ref{g_func}) for $b=0.1$ (a) and $b=0.9$ (b). The horizontal
  axis are colored black and the tangent points are denoted as black
  dots. }
\label{g_func_curves}
\end{center}
\end{figure}

Numerically solving Eq.~(\ref{g_func}), we find that either (i) it has
only two solutions for any value of $\lambda$ [see
  Fig.~\ref{g_func_curves}(a)], or (ii) it has either one or three
solutions for different values of $\lambda$ [see
  Fig.~\ref{g_func_curves}(b)]. When (i) occurs, the trivial solution of
Eq.~(\ref{theta_infty}) is $\theta(\infty)=1$ and there is no global
behavior adoption. When global behavior occurs, Eq.~(\ref{theta_infty})
has a non-trivial solution $\theta(\infty)<1$. At the critical point,
the equation
\begin{equation} \label{g_func}
\begin{array}{l}
g[\theta (\infty ),b,\gamma,\lambda ] = \sum\limits_{k' = 1}^{N - 1}
{\frac{{k'P(k')}}{{\langle k\rangle }}} \Theta (k',\infty )+
\frac{{\gamma [1 - \theta (\infty )](1 - \lambda )}}{\lambda } - \theta
(\infty )
\end{array}
\end{equation}
is tangent to the horizontal axis at $\theta(\infty)=1$.  Thus we find
the critical condition of the general social contagion model to be
\begin{equation} \label{d_g_d_theta}
\frac{{dg}}{{d\theta (\infty )}}{|_{{\theta}(\infty )=1}} = 0.
\end{equation}
Using Eq.~(\ref{d_g_d_theta}) we find the continuous critical
information transmission probability to be
\begin{equation} \label{lambda_c}
{\lambda _c^{\rm II}} = \frac{\gamma }{{\Gamma   + \gamma  - 1}},
\end{equation}
where
\[
\Gamma= \sum\limits_{k'} {\frac{{k'P(k')}}{{\langle k\rangle }}} (k' -
1)h(\frac{1}{{k'}},b).
\]
Numerically solving Eqs.~(\ref{theta_infty})--(\ref{lambda_c}), we find
$\lambda_c^{\rm II}$ to be a given adoption probability $h(x,b)$.  Here
$\lambda_c^{\rm II}$ is associated with adoption probability $h(x,b)$,
recovery probability $\gamma$, degree distribution $P(k)$, and average
degree $\langle k\rangle$.

In the second scenario, Eq.~(\ref{theta_infty}) can have three
solutions, and a saddle-node bifurcation can occur [see
  Fig.~\ref{g_func_curves}(b)].  Only the largest solution is valid
because only that value can be achieved physically.  Otherwise the fixed
point is the valid solution.
Changing $\lambda$ causes the physically meaningful stable solution of
$\theta(\infty)$ to jump to an alternate value. A discontinuous growth
pattern of $R(\infty)$ with $\lambda$ emerges, and solving
Eqs.~(\ref{theta_infty})--(\ref{lambda_c}) gives us the critical
transmission probability $\lambda^{\rm II}_c$ at which the discontinuity
occurs. When $b=0.9$, for different values of $\lambda$ the function
$g[\theta (\infty ),b,\gamma,\lambda ]$ is tangent to the horizontal
axis at $\lambda^{\rm II}_c=0.91$. When $\lambda < \lambda^{\rm II}_c$,
if there are three fixed points in Eq.~(\ref{theta_infty}), e.g.,
$\lambda=0.7$, the largest is the solution. When $\lambda = \lambda^{\rm
  II}_c$, the tangent point is the solution. When $\lambda >
\lambda^{\rm II}_c$, e.g., $\lambda=0.98$, the only fixed point is the
solution of Eq.~(\ref{theta_infty}), which abruptly drops to a small
value from a large value at $\lambda = \lambda^{\rm II}_c$ and causes a
discontinuous change in $R(\infty)$.

\begin{figure}[t]
\begin{center}
\includegraphics[height=12cm,width=15cm]{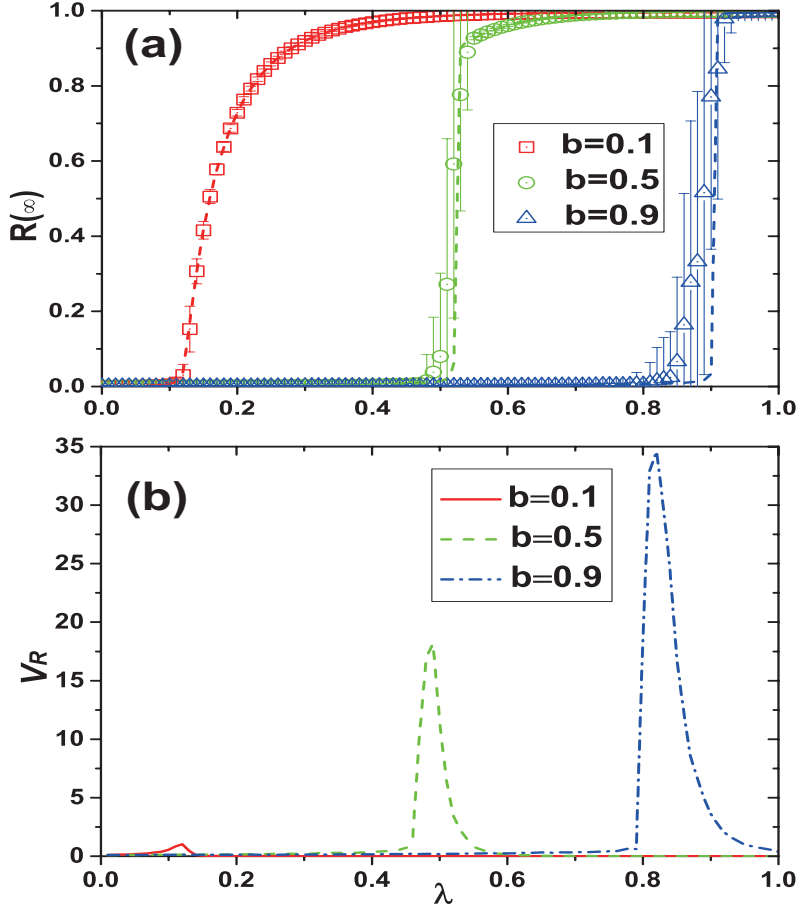}
\caption{(Color online) Illustration of effects of dynamical parameter
  $b$ with arbitrary $\lambda$, where $b$ is the LTI capacity parameter.
  (a) Under different $b$, the increase manners of final adoption size
  change from continuity at small $b$ (e.g. $b=0.1$) to discontinuity at
  large $b$ (e.g. $b=0.5$), which embody the second-order and
  first-order phase transition.  (b) $v_R$ numerically exhibits the
  fluctuation of $R(\infty)$ to intuitively emphasize the critical
  $\lambda_c^{\rm I}$ corresponding to the peak. The higher peak, the more
  abrupt the discontinuity of $R(\infty)$ is (see $b=0.1, 0.5 and
  0.9$).}
\label{R_VR_lambda}
\end{center}
\end{figure}

For a given $P(k)$, $\lambda$, and $\gamma$ and using the analytical method
similar to Eq.~(\ref{lambda_c}), we set $f(b)=\Gamma$ to be
\begin{equation} \label{fb_Gamma}
f(b)= \sum\limits_{k'} {\frac{{k'P(k')}}{{\langle k\rangle }}} (k' - 1)h(\frac{1}{{k'}},b),
\end{equation}
and
\begin{equation} \label{fb_frac}
f(b)= \frac{\gamma+\lambda-\gamma\lambda}{\lambda}.
\end{equation}
Using Eqs.~(\ref{fb_Gamma}) and (\ref{fb_frac}) gives us the critical
$b$ solution
\begin{equation} \label{b_c_2}
b_c^{\rm II} = f^{-1}(\frac{\gamma+\lambda-\gamma\lambda}{\lambda}).
\end{equation}
From this theoretical analysis and using non-redundant memory, the
social contagion with LTI character displays first and second-order
phase transitions.

\begin{figure}[t]
\begin{center}
\includegraphics[height=12cm,width=10cm]{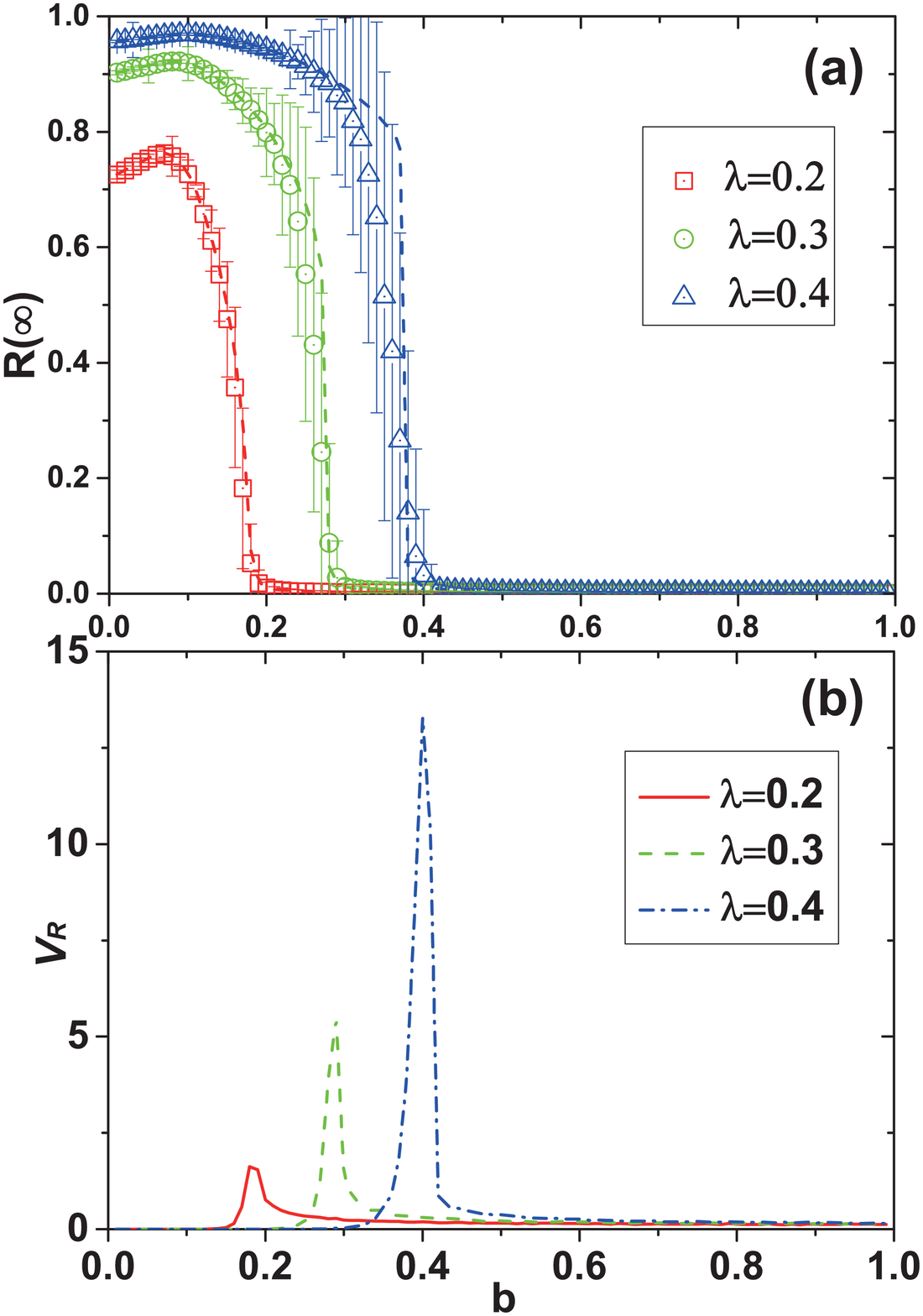}
\caption{(Color online) Illustration of effects of dynamical parameter
  $\lambda$ with arbitrary $b$, where $\lambda$ is behavior transmission
  probability and $b$ is LTI capacity parameter.  (a) Under different
  $\lambda$, the change pattern of $R(\infty)$ with $b$ shows a first
  increase and then a decrease.  Ultimately, the $R(\infty)$ vanishes to
  zero.  (b) To find the critical point where the $R(\infty)$ vanishes,
  $v_R$ is introduced here. Besides, the peaks of $v_R$ lines correspond
  to critical points, and the higher the peaks the sharper the jump to
  zero.}
\label{R_VR_b}
\end{center}
\end{figure}

\section{Numerical simulations}
Extensive experiments have been performed on ER and SF, where the
network size, mean degree, and recovered probability are $N=10^4$,
${\langle k\rangle }=10$, and $\gamma=1.0$, respectively. The relative
variance $v_R$ is designed numerically to determine the size-dependent
critical values $\lambda_c^{\rm II}$ and $b_c^{\rm II}$. The relative variance
of $R(\infty)$\cite{Chen2014} is defined as 
\begin{equation}\label{V_R}
{v_R} = \frac{{\langle {{(R(\infty ) - \langle R(\infty )\rangle
        )}^2}\rangle }}{{{{\langle R(\infty )\rangle }^2}}},
\end{equation}
where $\langle...\rangle$ is the ensemble average.  The value of $v_R$
shows the peaks (indicating phase transitions) of $R(\infty)$ when a
dynamical parameter is varied.  Thus we know that the $\lambda_c^{\rm II}$
and $b_c^{\rm II}$ correspond to the maximum $v_R$ under different values of
$\lambda$ and $b$, respectively.

To study social contagions on ER networks, we examine the final behavior
adoption size $R(\infty)$ as a function of the transmission probability
$\lambda$ for different values of the LTI capacity parameter $b$ when
$\gamma = 1.0$. Figure~\ref{R_VR_lambda}(a) shows that a bifurcation
analysis of Eq.~(\ref{theta_infty}) reveals that the LTI on adopted
informants at different $b$ affects the type of phase transition. When
the LTI capacity is strong, e.g., $b=0.1$, the system exhibits a
second-order phase transition because a small $b$ value indicates that a
low ratio of informants can cause massive behavior adoptions even when
the transmission rate $\lambda$ is low. When the LTI capacity is weak,
e.g., $b=0.5$ or 0.9, the system exhibits a first-order phase transition
because a high $b$ value indicates that a high ratio of informants and
low transmission rate $\lambda$ with a low probability of transmitting
information does not substantially increase the informant ratio of
susceptible individuals. When $b$ is high, the transmission rate
$\lambda$ must exceed a critical point for there to be a massive
information reception by many individuals that greatly increases the
informant ratio of susceptible individuals, which results in an abrupt
outbreak of behavior adoption.

\begin{figure}[t]
\begin{center}
\includegraphics[height=14cm,width=12cm]{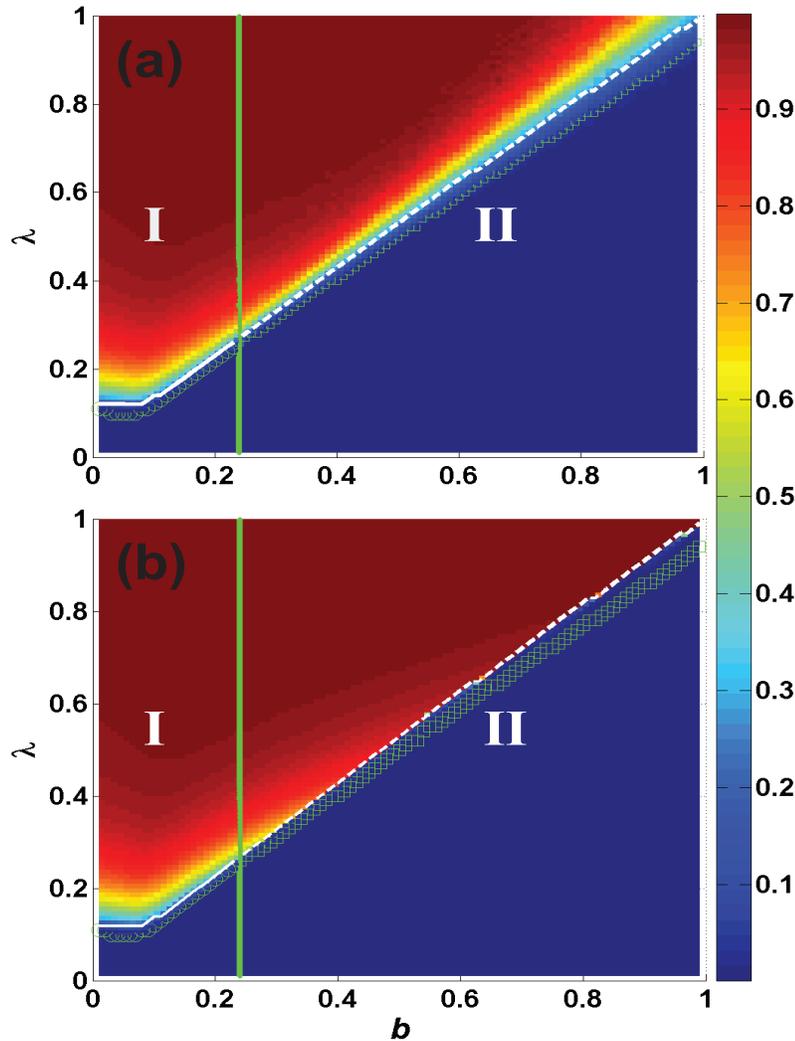}
\caption{Dependance of $R(\infty)$ on $b$ and $\lambda$ on ER network.
  Color-coded values of $R(\infty)$ are obtained from numerical
  simulations (a) and theoretical solutions (b). Herein, the theoretical
  solutions are achieved through Eqs.~(\ref{active_a_k})-(\ref{S_t})
  and~(\ref{d_theta_t})-(\ref{d_R_t}).  The parameter plain is divided
  into two regions by $b^*$ which is obtained from
  Eqs.~(\ref{d_g_d_theta}),~(\ref{lambda_c}) and~(\ref{b_c_2}).  In
  regions $I$, $R(\infty)$ shows a continuous increase and undergoes a
  second-order phase transition.  In contrast, $R(\infty)$ exhibits a
  discontinuous increase and undergoes a first-order phase transition in
  region $II$.  The solid white line from theoretical method and green
  circles from numerical simulation all represent the critical
  $\lambda_c^{\rm II}$ in region $I$.  And the dashed white line from
  theoretical method and green rectangles from numerical simulation as
  well denote the the critical $\lambda_c^{\rm II}$ in region $II$.}
\label{lambda_b_plain}
\end{center}
\end{figure}

We calculate the theoretical value of $\lambda_c^{\rm II}$ using
Eqs.~(\ref{theta_infty})--(\ref{d_g_d_theta}).  To locate the numerical
critical points, we examine $v_R$ shown in
Fig.~\ref{R_VR_lambda}(b). Our theory coincides with these simulation
results well, except when $\lambda$ is close to the critical information
transmission probability.  The deviations between our predictions and
the simulations are caused by the finite-size effects of networks, and
the strong dynamical correlations among the states of neighbors.

Figure~\ref{R_VR_b} shows an analysis of $R(\infty)$ versus $b$ for
different $\lambda$ values.  For a given $\lambda$, $R(\infty)$ changes
nonmonotonically with $\lambda$. In particular, $R(\infty)$ first
increases with $b$ and then decreases discontinuously to zero. Thus
there is an optimal $b_o$ value at which $R(\infty)$ reaches its maximum
value. To explore the reason, given $\lambda$,
when $b$ is smaller than the optimal $b$,
LTI capacity is very strong and a lot of neighbors around the susceptible turn into informed state,
leading to ever-increasing $R(\infty)$. When $b$ becomes greater,
LTI capacity gradually decreases showing reduced capacity in informing neighbors,
but $R(\infty)$ still keeps increasing until $b$ exceeds the optimal $b$.
At the optimal $b$, the informing process enters in a balance status and the $R(\infty)$ also reaches the maximum.
Gradually, when $b$ is greater than the optimal $b$, LTI capacity and informing capacity considerably degrade,
leading to insufficient informed neighbors to effectively support further informing,
so $R(\infty)$ enters in ever-declining status until zero.
The critical point can be located by
studying $v_R$ as shown Fig.~\ref{R_VR_b}(b). Again,
our theory agree well with the numerical simulations.

Figure~\ref{lambda_b_plain} shows a study of the phase transition plane
$(\lambda,b)$. According to the type of phase transition, the parameter
plane $(\lambda, b)$ is divided into two regions by the critical value
of $b$ ($b^*=0.237$), which can be obtained using Eqs.~(\ref{g_func})
and (\ref{lambda_c}).  In region I, i.e., $(b \le b^*)$, $R(\infty)$
increases continuously and exhibits a second-order phase transition.  In
region II, i.e., $(b>b^*)$, $R(\infty)$ increases discontinuously with
$\lambda$ and exhibits a first-order phase transition.  There is also a
crossover in the phase transition. The numerical simulation generally
agrees with theoretical solution.

\begin{figure}[t]
\begin{center}
\includegraphics[height=14cm,width=14cm]{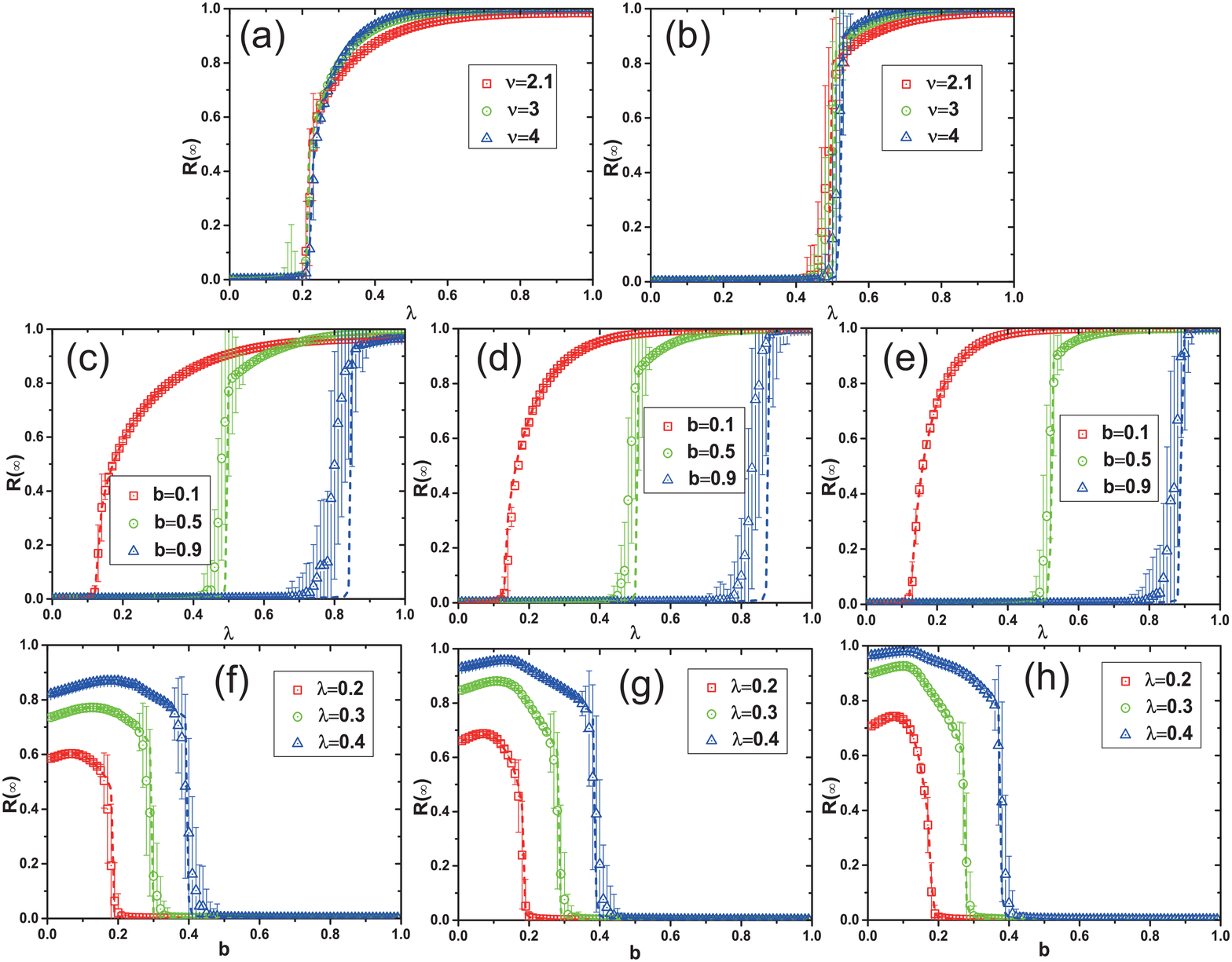}
\caption{(Color online) Effect of network heterogeneity in social
  contagion dynamics.  For scale-free network with mean degree $\langle
  k\rangle=10$ and network size $N=10000$, dependence of $R(\infty)$ on
  $b$, $v$, and $\lambda$ is explored under parameter values.  Subgraph
  (a) and (b) demonstrate the impact on $R(\infty)$ of separate degree
  exponent $v$, respectively under $b=0.2$ and $0.5$. Then, subgraph
  (c), (d), and (e) exhibit the results of $b$ influencing $R(\infty)$,
  referring to degree exponent $v=2.1$, $3$, and $4$. Furthermore, for
  rigorousness, subgraph (f), (g), and (h) proceed to show the changes
  of $R(\infty)$ based on different transmission probability $\lambda$,
  also separately under $v=2.1$, $3$, and $4$. As expected, the
  theoretical solutions, denoted by dash line, perfectly coincide with
  numerical values, marked by symbols.}
\label{SF_topology}
\end{center}
\end{figure}

Figure~\ref{SF_topology} shows a study of the effects of the
heterogeneity of degree distribution on social contagion. Here we focus
on the SF network with different degree exponents $v$. We set the
average degree and network size to be $\langle k\rangle=10$ and
$N=10^4$, respectively. Figures~\ref{SF_topology}(a) and
\ref{SF_topology}(b) show that the heterogeneity of degree distribution
does not change the type of phase transition when $b=0.2$ and $0.5$,
respectively.

Figure~\ref{SF_topology}(a) shows that when $b=0.2$ the increase of
$R(\infty)$ exhibits a change from a discontinuous first-order to a
continuous second-order phase transition, when $v$ rises from $2.1$ to
$4$. Figure~\ref{SF_topology}(b) shows, in contrast, when $b=0.5$,
$R(\infty)$ increases and exhibits the same pattern of first-order phase
transition at any $v$ value and jumps higher at the critical
$\lambda_c^{\rm II}$ when $v=2.1$, $3$, and
$4$. Figures~\ref{SF_topology}(c)--\ref{SF_topology}(e) show that when
$v=2.1$, $v=3$, and $v=4$ increasing $b$ also changes the growth pattern
of $R(\infty)$ from a second-order phase transition to a first-order,
but that the final adoption size increases with $v$, i.e., the
heterogeneous degree distribution does not impede the change in phase
transition.  Figures~\ref{SF_topology}(f)--\ref{SF_topology}(h) show
that when $v=2.1$, $v=3$, and $v=4$ the transmission probability
$\lambda$ influences the final range of $R(\infty)$ that increases the
number of individuals when $v$ is higher.  For all three values and
under each $\lambda$ the optimal LTI capacity parameter $b$ that
maximizes $R(\infty)$ appears, and the critical LTI capacity point of
$b_c^{\rm II}$ reduces $R(\infty)$ to zero, even though a higher value of $v$
promotes a wider spreading of behavior information. A heterogeneous
degree distribution always causes a change of phase transition as in
(c)-(e) and of optimal and critical LTI capacity parameters as in
(f)-(h). In addition, the theoretical solutions (dashed lines) agree
with the numerical values (symbols) in all subsections of
Fig.~\ref{SF_topology}.

\section{Conclusions}
The local trend imitation (LTI) phenomenon is ubiquitous and strongly
affects the dynamics of social contagions. We have proposed a social
contagion model that uses a tent-like adoption function to
systematically study the role of LTI. We use an edge-based compartmental
theory to describe the model and find that the theoretical
predictions agree with the numerical simulations. We also perform
extensive numerical simulations on ER networks. We find that when the
LTI capacity is weak the size of the final behavior adoption grows
discontinuously, i.e., the system exhibits a first-order transition, but
when the LTI capacity is strong the size of the final behavior adoption
grows continuously, i.e., the system exhibits a second-order phase
transition.  Thus there is a crossover in the phase transition type. For
a given probability of information transmission, there an optimal LTI
capacity at which the final behavior adoption size is markedly
increased. We also find that degree heterogeneity does not qualitatively
alter these phenomena.

\section*{Acknowledgments}
This work was supported by the National Natural Science Foundation of
China (Nos. 61602048, 61673086, 61673085) and the Fundamental Research Funds for
the Central Universities.

\end{document}